\documentclass[a4paper,conference]{IEEEtran}
\usepackage{cite}

\usepackage[pdftex]{graphicx}
\graphicspath{{./figures/}}

\usepackage[cmex10]{amsmath}
\usepackage{array}

\usepackage{mdwmath}
\usepackage{mdwtab}
\usepackage[caption=false,font=footnotesize]{subfig}
\usepackage{url}

\usepackage[usenames,dvipsnames]{color}
\usepackage{epstopdf,gensymb,xfrac,fixltx2e,listings}

\usepackage[bookmarks,pdfdisplaydoctitle,pdftitle={Metadata for Energy Disaggregation},hidelinks,colorlinks,linkcolor=black,citecolor=black,urlcolor=black]{hyperref}

  { \raggedright 
    \setlength{\parskip}{0.4em}}
  {}

\begin{document}
%
\title{Metadata for Energy Disaggregation}


\author{
\IEEEauthorblockN{Jack Kelly and William Knottenbelt}
\IEEEauthorblockA{Department of Computing, Imperial College London, UK.\\
e-mail: jack.kelly@imperial.ac.uk}
}


\maketitle

\begin{abstract}

Energy disaggregation is the process of estimating the energy consumed
by individual electrical appliances given only a time series of the
whole-home power demand.  Energy disaggregation researchers require
datasets of the power demand from individual appliances and the
whole-home power demand.  Multiple such datasets have been released
over the last few years but provide metadata in a disparate array of
formats including CSV files and plain-text README files.  At best, the
lack of a standard metadata schema makes it unnecessarily
time-consuming to write software to process multiple datasets and, at
worse, the lack of a standard means that crucial information is simply
absent from some datasets.  We propose a metadata schema for
representing appliances, meters, buildings, datasets, prior knowledge
about appliances and appliance models.  The schema is relational and
provides a simple but powerful inheritance mechanism.

\end{abstract}
%
\IEEEpeerreviewmaketitle

\section{Introduction}

Research suggests that consumers are better able to reduce their
energy consumption if given an itemised, appliance-by-appliance energy
bill rather than a bill which only describes aggregate
consumption~\cite{Fischer2008}.  Energy disaggregation is the process
of estimating the energy consumed by individual appliances in a home
given a time series of the whole-home power demand.  A typical use case
would be to provide consumers with an estimated itemised electricity
bill without requiring the expense of installing separate meters on
every appliance.

Research into energy disaggregation (also known as `non-intrusive load
monitoring' or NILM) began with George Hart's pioneering work in
1984~\cite{Hart1984,Hart1992}.  A recent resurgence has been triggered
by a combination of high energy bills and the introduction of `smart
electricity meters'.  In the quest to design and implement a high
performance energy disaggregation system, researchers require several
types of data:

The primary requirement is for datasets which record the power demand
of whole homes as well as the `ground truth' power demand of
individual appliances within the home.  In 2011, researchers at MIT
released the first public dataset for energy disaggregation
research~\cite{REDD}.  Since then, over ten more datasets have been
released~\cite{tracebase, PecanStreet, BLUED, IAWE, smart, HES} and,
in March 2014, a project called Wiki
Energy\footnote{\url{http://wiki-energy.org/}} launched to share
datasets online.  These datasets have been well received but, because
each dataset uses a different file format, it is time consuming to
import multiple datasets.  This is an issue because an important criteria
for evaluating any machine learning algorithm is how well it
generalises across multiple datasets. A further challenge with
existing datasets is that machine-readable metadata is often minimal
and uses a schema and vocabulary unique to that dataset.  At best, the
lack of a standard metadata schema makes it time-consuming to write
software to process multiple datasets.  At worse, some datasets simply
lack sufficient metadata to allow the data to be properly interpreted.
For example, the mains wiring connecting meters to each other and to
appliances forms a tree (with the whole-house meter at the root and
appliances at the leaves).  In some datasets, this tree structure has
more than two levels (i.e. when an appliance turns on or off, the
resulting change in power consumption is sensed by more than two
meters) yet the metadata rarely specifies the wiring tree.

A secondary requirement is for data describing the behaviour of
appliances (e.g. a probability distribution describing the typical
times per day that each appliance is used).  This prior knowledge can
be used to fine-tune the estimates produced by a disaggregation
system.  Such data is currently available in research papers and
industry reports but not in a machine-readable form.

Finally, consumers are unlikely to put effort into training a
disaggregation tool.  As such, if open-source disaggregation solutions
such as NILMTK\cite{NILMTK} are to be
viable as consumer-facing disaggregation solutions then researchers
must distribute pre-trained models for each appliance.  If these
models adhere to a standard metadata schema then multiple software
systems can exchange models.

Against this background, we propose the first draft of a hierarchical
metadata schema for energy disaggregation.  Specifically, our schema
models electricity meters, appliances (including prior knowledge such
as probability distributions describing typical times of use and
parameters describing inferred models of appliances), buildings and
datasets.

Although we have made every effort to ensure that our proposed schema
and controlled vocabularies capture the information present in all the
datasets we are aware of, our schema can undoubtedly be improved and
so the schema is presented as an open-source
project\footnote{\url{https://github.com/nilmtk/nilm_metadata}} (under a
permissive Apache 2.0 license) to which contributions are most
welcome!

In this paper we first outline related work; then we describe the
design of our metadata schema; then look at the implementation of a
schema validator and finally we discuss conclusions and future work.

\section{Related work}

The general principal of using metadata to describe research datasets
is not new.  For example, the not-for-profit organisation
DataCite\footnote{\url{http://www.datacite.org}} (established in 2009)
publishes the DataCite Metadata Schema for describing research
datasets.

In late-2013, the Nature Publishing Group launched a journal called
`Scientific Data`\footnote{\url{http://www.nature.com/scientificdata}}
for describing datasets.  The machine-readable description of each
dataset is captured using the ISA\_Tab metadata
specification\footnote{\url{http://isa-tools.org}} which specifies a
hierarchical schema consisting of the `investigation' (the project
context), the `study' (a unit of research) and the `assay' (an
analytical measurement).

Biology researchers have embraced the need for metadata schemas and
controlled vocabularies as demonstrated by, for example, The Open
Biological and Biomedical Ontologies
database\footnote{\url{http://www.obofoundry.org}} which aims to
enable the creation of a suite of interoperable reference ontologies
for the biomedical domain.

Few metadata projects are specifically for energy datasets.  One
notable project is Project
Haystack\footnote{\url{http://project-haystack.org}} which is an open
source initiative to develop taxonomies and tagging conventions for
building equipment and operational data.  Amongst other achievements,
Haystack defines a language for describing electricity meters,
including which parameters each meter records and the relationships
between meters and between meters and loads.  But Haystack is
primarily targeted at large commercial buildings rather than domestic
buildings, and does not define a controlled vocabulary for appliance
names, let alone more granular detail about appliances.

The UK Energy Research Council's Energy Data Centre provides a simple
schema\footnote{\url{http://ukedc.rl.ac.uk/format.html}} based
on the Dublin Core Metadata
Initiative\footnote{\url{http://dublincore.org}} (DCMI).

The Power Consumption
Database\footnote{\url{http://www.tpcdb.com}} is a community project
which aims to collect a database of appliance power consumption information.

To summarise the related work: there are many metadata projects for
describing research datasets in general but only a small number of
metadata projects for describing energy datasets.  To the best of our
knowledge, there are no existing metadata schemas specifically for
describing objects relevant to energy disaggregation.  Existing
datasets for energy disaggregation do provide some metadata (e.g. a
text file mapping appliance names to recording channels) but this
metadata does not use a controlled vocabulary and often provides scant
details.

\section{Design}

The NILM Metadata schema models several objects relevant to energy
disaggregation: electricity meters, appliances, prior knowledge about
appliances, appliance models, buildings and datasets. The schema
specifies property names for each object, the type for each value and
controlled vocabularies (e.g. for appliance names and categories).

\begin{figure}
  \centering
  \includegraphics{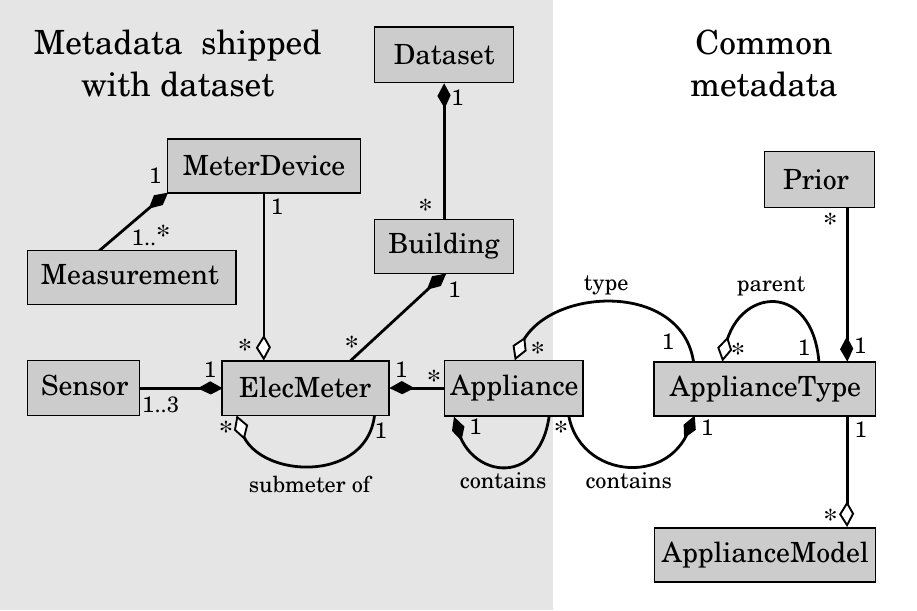} 
  \caption{UML Class Diagram showing the relationships between
    classes.  A dark black diamond indicates a `composition'
    relationship whilst a hollow diamond indicates an
    `aggregation'. For example, the relationship between `Dataset' and
    `Building' is read as `\textit{each Dataset contains any number of
      Buildings and each Building belongs to exactly one Dataset}'.
    We use hollow diamonds to mean that objects of one class refer to
    objects in another class.  For example, each Appliance object
    refers to exactly one ApplianceType.  Instances of the classes in
    the shaded area on the left are intended to be shipped with each
    dataset whilst objects of the classes on the right are common to
    all datasets and are stored within the NILM Metadata project.
    Some ApplianceTypes contain Appliances, hence the box representing
    the Appliance class slightly protrudes into the `common metadata'
    area on the right.}
  \label{fig:schema}
\end{figure}

\begin{figure*}
  \centering
  \includegraphics{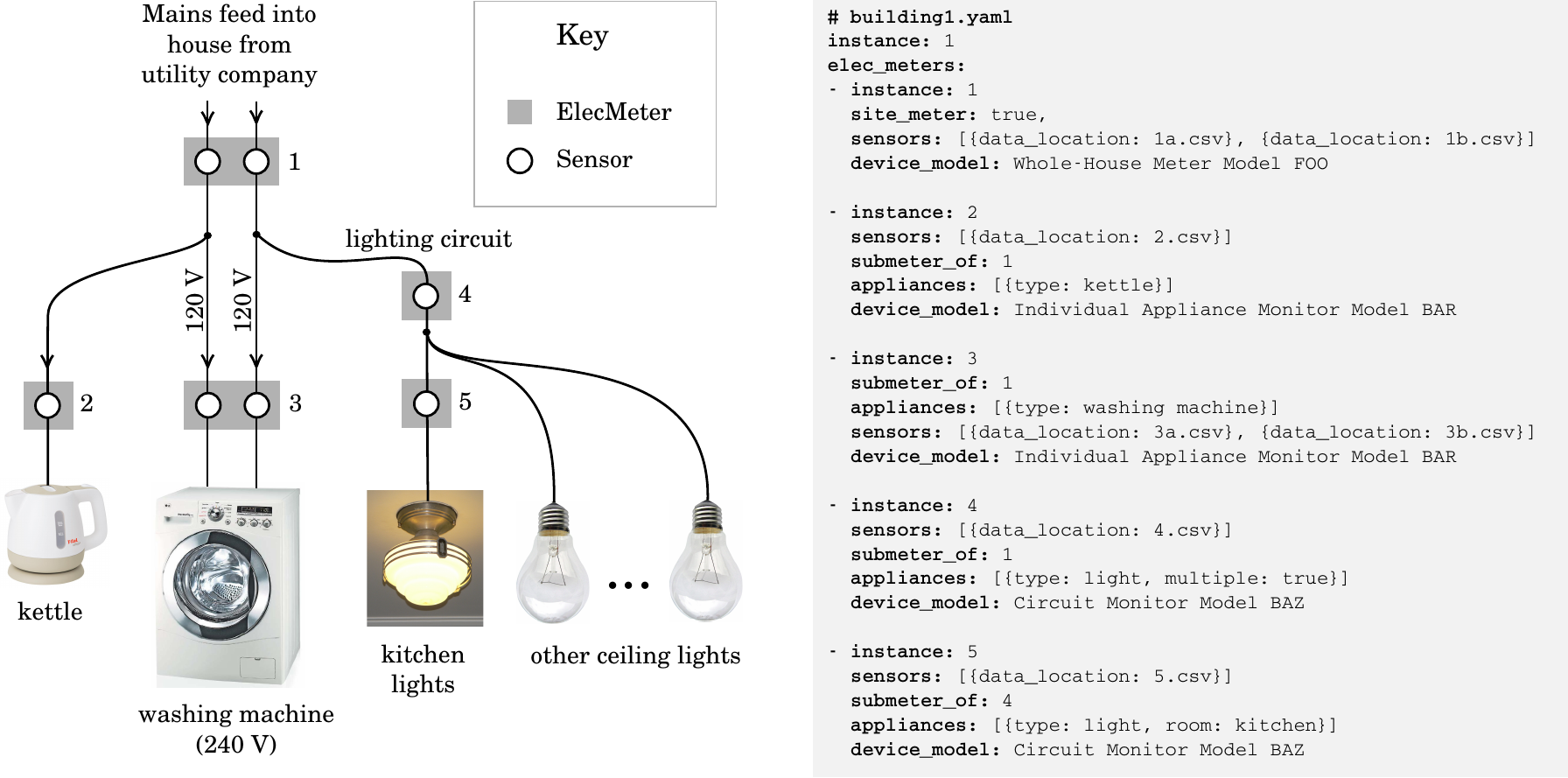} 
  \caption{The illustration on the left shows a cartoon mains wiring
    diagram for a domestic building. Black lines indicate mains
    wires. This home has a split-phase mains supply (common in North
    America, for example). The washing machine draws power across both
    splits.  All other appliances draw power from a single split.  The
    text on the right shows a minimalistic description (using the NILM
    Metadata schema) of the wiring diagram on the left.}
  \label{fig:circuit}
\end{figure*}

A UML Class diagram showing the relationship between classes is shown
in Figure~\ref{fig:schema} and a brief example metadata instance is
shown in Figure~\ref{fig:circuit}.

In the sections below, we describe our `dataset' and `building'
schemas; the distinction between \emph{meters} and \emph{appliances};
the representation of electricity meters; the representation of the
mains wiring; the inheritance mechanism for appliances;
categorisation; the containment mechanism that allows an appliance to
contain other appliances; prior knowledge; and finally our
representation of learnt models.

\subsection{ Dataset}
 
NILM Metadata places the primary objects of interest into a tree
shaped hierarchy (Figure~\ref{fig:schema}).  At the root is a dataset
object. This contains buildings, each of which contains electricity
meters, many of which measure the power demand of appliances.

This tree hierarchy captures all datasets we are aware of except one:
the `tracebase' dataset~\cite{tracebase} describes appliances without
their building context.  To handle tracebase, meter objects in NILM
Metadata can be directly contained within a dataset object (without
requiring a building object).

\begin{flushleft}
The dataset schema records properties such as
\texttt{publication\_date$^*$, rights\_list$^*$, geospatial\_coverage$^*$,
  temporal\_coverage$^*$, funding, creators$^*$, related\_documents$^*$, timezone}
and \texttt{geo\_location} ($^*$ the starred properties are adapted
from Dublin Core.)
\end{flushleft}

\subsection{ Building}
Buildings are identified by an integer property \texttt{instance}
(unique within the dataset). Each building may have a list of rooms
(using a controlled vocabulary for room names), and some properties
shared with dataset: \texttt{temporal\_coverage, geo\_location,
  timezone}.  These properties default to the values set in the parent
dataset but can be overridden per building.  Each building contains a
\texttt{elec\_meters} property which stores a list of ElecMeter objects.

\subsection{ Meters are distinct from appliances}
A tempting simplification would be to assume a one-to-one relationship
between electricity meters and appliances.  But we often observe
one-to-\emph{many} relationships (e.g. multiple appliances plugged
into a multi-way mains adapter which, in turn, is connected to a
single meter) and we occasionally observe many-to-one relationships
(e.g. in the US and Canada many large domestic appliances like washing
machines draw a total of 240 volts from two 120 volt `split-phase'
supplies found in a typical house and some datasets use two meters per
240 volt appliance).  We frequently observe situations where some
appliances are not submetered.  To handle the case where a single
appliance receives more than one power supply (e.g. split-phase or
three-phase power), we allow each ElecMeter object to contain between one
and three Sensor objects.  Each Sensor models the physical sensors
recording data in the field.  To handle the case where a single meter
is connected to multiple appliances, each ElecMeter can contain any number
of Appliance objects.  The meter may also specify a
\texttt{dominant\_appliance} property to specify if a single appliance
is on more often than other appliances on that meter.

\subsection{ ElecMeters and MeterDevices}
Each MeterDevice object records properties which apply to a specific
model of electricity meter.  For example, the \texttt{sample\_period}
in seconds, the \texttt{measurements} recorded by the meter
(e.g. voltage, reactive power, active energy etc), the meter 
\texttt{manufacturer} and \texttt{model}.

ElecMeter objects represent each physical meter installed in a
building. Each ElecMeter references exactly one
\texttt{device\_model}.  ElecMeter is also the place where any
pre-processing carried out on the data can be described (for example,
have gaps been filled? Or unrealistic values been removed?)

\subsection{ Mains wiring}
Each building in a typical dataset will have one meter which records
the aggregate, whole-building mains power demand.  Downstream of this
meter might be meters which measure entire circuits within the
building (e.g. the lighting circuit).  Finally, there are often meters
which measure individual appliances.  An example wiring diagram is
shown in Figure~\ref{fig:circuit}.

As such, the mains wiring connecting meters with each other can be
described as a tree.  Each ElecMeter can specify either a
\texttt{submeter\_of} property (the numeric ID of the upstream meter)
or a \texttt{site\_meter} property (a boolean flag which is set to
\texttt{true} if this meter measures the whole-building aggregate).
The property names `\texttt{submeter\_of}' and `\texttt{site\_meter}'
are adapted from Project Haystack.  The wiring hierarchy can be any
depth.  In large, commercial installations, a meter in one building
may be downstream of a meter in another building.  This case can be
handled by specifying the numeric ID of the other building using the
\texttt{upstream\_meter\_in\_building} property.  If this property
absent then we assume the upstream meter is in the same building.

\subsection{ Appliance and ApplianceType}
With each dataset, we specify a set of Appliance objects.  Each
Appliance object represents an appliance \textit{instance} in the
dataset. Each Appliance object in a dataset has a \texttt{type}
property which refers to an ApplianceType object.  ApplianceType objects are
not shipped with the dataset; instead they are stored within NILM
Metadata and embody the controlled vocabulary of appliance names and
all the prior knowledge about appliance types (e.g. the categories
each appliance type falls within, probability distributions describing
the power demand for the appliance etc).

\subsection{ Inheritance for ApplianceTypes}
Electrical appliances can be described as a hierarchical tree of
objects.  For example, a `wine cooler' can be considered a
specialisation of a `fridge' and, as such, inherits properties from
fridges.  

Inheritance is a well-established technique in software engineering
for maximising code re-use.  NILM Metadata implements a simple but
powerful form of inheritance known as prototype-based inheritance
(first implemented in the Self programming language
~\cite{Chambers1989} and used in JavaScript). Objects in
prototype-based programming languages are not instances of a class
but, instead, inherit from any other object (the `parent' or
`prototype' object).  In NILM Metadata, each ApplianceType object has a
`parent' from which it inherits properties.  These properties can be
modified by the child and the child can specify properties not
specified by the parent.  The inheritance tree can be any depth.

Inheritance follows a small number of rules.  If a property is
contained in the parent and absent in the child then it is copied to
the child.  If a property is present in both parent and child then it
is handled differently depending on the type of the property:

\begin{enumerate}
  \item \textbf{list (array)} objects become the union of the parent
    and child lists.
  \item \textbf{scalar} objects in the child override (`shadow')
    properties in the parent.
  \item \textbf{objects} (dictionaries) are recursively updated using
    the rules above.
\end{enumerate}

Child objects can specify a \texttt{do\_not\_inherit} property (a list
of property names) to avoid inheriting named properties.

\noindent
\textbf{Subtypes versus inheritance}. Appliance objects have a
\texttt{subtype} property (which must be set to a member of the
appliance type's \texttt{subtypes} set).  What is the difference between a
subtype and a child object?  Subtypes are useful when two related
appliances are so similar that we can safely ignore the differences
for the purposes of energy disaggregation.  For example, an analogue
radio and a digital radio are sufficiently similar to mean that they
can both be subtypes of the `radio' object.  On the other hand, an
electric cooker has a significantly different electricity load profile
compared to a cooker fuelled by natural gas, so these are separate
objects.

\noindent
\textbf{Additional properties}. Some appliances have rare properties.
For example, a television might have a \texttt{screen\_size} property.
We do not want to pollute the common `appliance' schema with these
properties (because, for example, it makes no sense for a cooker to be
able to specify a \texttt{screen\_size} property!).  Instead,
appliance objects can define an \texttt{additional\_properties}
property.  This is specifies the schema for any additional properties
(using JSON Schema).  \texttt{additional\_properties} is inherited
using the same rules as any other property.

\subsection{ Appliance categorisation}
When analysing domestic power consumption, we often want to group
appliances into certain categories.  For example, we might want to ask
`what is the total energy consumption for all consumer electronics?'.

Domestic appliances are traditionally classified as one of `wet',
`cold', `consumer electronics', `ICT', `cooking', `lighting' or
`heating'.

An alternative classification is a simple binary
classification of `large appliances' (e.g. dish washer) versus `small
appliances' (e.g. a radio).
 
A more finely-grained classification based on the electrical
properties of appliances was proposed by \mbox{Tsagarakis \textit{et
    al}~\cite{Tsagarakis2013a}}.  For example, the taxonomy proposed
by Tsagarakis \textit{et al.} splits lighting into general incandescent lamps,
fluorescent lamps and light-emitting diode (LED) sources.  An
appliance can have multiple classifications from this taxonomy.

Yet another taxonomy for domestic appliances is the Google product
taxonomy\footnote{\url{https://support.google.com/merchants/answer/160081}}
(used on Google Shopping).  This taxonomy is a tree which we represent
as list of classifications.

NILM Metadata currently supports all four taxonomies listed above and
it would be trivial to add more.  We specify a controlled vocabulary
for the category names.  Our appliance schema specifies a
\texttt{categories} property which is an object with the following
properties: \texttt{traditional} (\emph{string}), \texttt{size}
(\emph{string}), \texttt{electrical} (\emph{array of strings}) and
\texttt{google\_shopping} (\emph{array of strings}).  At present, all
ApplianceTypes the NILM Metadata have a `traditional' classification
and many have classifications for the other taxonomies.

\subsection{ Appliances can contain other appliances}
Some appliances can be modelled as a container of other objects. For
example, a washing machine can be modelled as a drum motor and a water
heating element (and a few other components).  Appliance (and
ApplianceType) objects in NILM Metadata have a \texttt{components}
property which stores an array of appliance objects.  Containment is
recursive and can be of any depth.

Of course, \emph{all} appliances can be decomposed into components.
Do we model each individual resistor and transistor? No; the end-goal
is to model appliances only in sufficient detail to allow an energy
disaggregation system to identify the whole appliance given prior
knowledge of the components.  As such, we only describe individual
components if their electrical behaviour is observable from a typical
mains electricity meter.  It is also important that components be
truly separate entities from an electrical perspective.  For example,
a fridge freezer should \emph{not} be modelled as containing both a
fridge and a freezer because that would imply that a fridge freezer
has two separate compressors but - as far as we are aware - fridge
freezers typically have one compressor.

If an appliance contains multiple instances of the same component then
we use the \texttt{count} property in the component to specify the
number of instances.  If an appliance contains multiple instances of
the same component but the exact number of components is unknown then
set \texttt{multiple} to `true'.

The container appliance inherits categories from each of its
components.  This is useful mostly for the `electrical' taxonomy.  For
example, if we model a washing machine as a motor and a heater
then the washing machine inherits the appropriate
electrical classifications from both the motor and the heater.

Our representation of lighting exploits NILM Metadata's containment
mechanism. We distinguish between the light fitting (also called the
luminaire or fixture) and the electric lamp(s) within each fitting.
We have a `light' object which contains any number of `lamps' (of
which there are several kinds including `LED lamp' and `incandescent
lamp'). Light objects can also contain a `dimmer' object.

\subsection{ Prior knowledge}

Prior knowledge can be exploited to improve disaggregation
performance.  Examples of prior knowledge include: the distribution of 
on-powers of an appliance; the typical time of use per day or per week;
correlations with other appliances (e.g. the computer monitor is often
on when the computer is on).

NILM Metadata specifies a `prior' object which holds several
properties, the two most important of which are
\texttt{distribution\_of\_data} (the distribution of the
data expressed as normalised frequencies per discrete bin (for
continuous variables) or per category (for categorical variables)) and
\texttt{model} (which describes a model fitted to describe the
probability density function (for continuous variables) or the
probability mass function (for discrete variables)).  We can also
specify the \texttt{source} of data (is it a subjective guess, or the
result of primary data analysis, or taken from a published paper?),
whether the prior is \texttt{specific\_to} a country and what
\texttt{training\_data} was used to generate the prior.

Each `appliance' object has a `distributions' property which is an
object with the following properties (each property is an array of
priors): \texttt{on\_power, on\_duration, off\_duration,
  usage\_hour\_per\_day, usage\_day\_per\_week,
  usage\_month\_per\_year, rooms, subtypes, appliance\_correlations,
  ownership, ownership\_per\_country, ownership\_per\_continent}

We store an \emph{array} of priors (rather than a single prior) for
each distribution.  This allows us to store multiple beliefs about
each distribution (which could be combined using Bayesian statistics).
For example, we might find several published papers which provide
evidence about the distribution over the power consumption of an
appliance.  Furthermore, NILM Metadata collects all relevant priors as
it descends the inheritance hierarchy for each object (for example, a
`wine cooler' might not have any priors associated with it but it will
inherit prior knowledge from its parent `fridge' object).  Of course,
priors from a distant ancestor are less relevant than priors from a
recent ancestor so, as we traverse the inheritance tree, we tag each
prior with a \texttt{distance} property (a positive integer indicating
the number of `generations' away the prior is from the appliance in
question).

\subsection{\;\; Learnt models of appliances}

End-users of domestic disaggregation software are unlikely to put any
effort into training the system.  This means that we must use either a
supervised learning algorithm with pre-trained models or an
unsupervised disaggregation algorithm (which must still have some form
of prior appliance model to be able to provide human-readable names
for each appliance).  As such, we specify a simple `appliance model'.
This has properties such as \texttt{model\_type} (a controlled
vocabulary with terms such as `HMM' for hidden Markov model),
\texttt{training\_data, date\_prepared} etc.  The model's parameters
are stored in a \texttt{parameters} object.

For each model type (e.g. `HMM') and for each appliance type
(e.g. `fridge'), our `appliance model' schema can be used to fully
specify an appliance model that has been learnt from the data.  This
allows the model parameters to be inferred and shared by a researcher
and then any user can take advantages of these models for
disaggregation.  The end result is that only a small number of
researchers need to put effort into generating models and then 
anyone with an internet connection can make use of the models in a
disaggregation system.

\section{Implementation}

The syntactic elements of the schema are specified using JSON Schema
Draft 4\footnote{\url{http://json-schema.org/}}. The code which
implements the semantics of NILM Metadata and performs validation is
written in Python.  We make use of the
\texttt{jsonschema}\footnote{\url{https://github.com/Julian/jsonschema}}
package for validation and
\texttt{PyYAML}\footnote{\url{http://pyyaml.org/wiki/PyYAML}} for
loading YAML files.  Metadata instances can be written in JSON or YAML.

Prior to validating each appliance, the \texttt{properties} object
specified by the `appliance' schema is updated with concatenated
\texttt{additional\_properties} specified by the appliance's ancestors.

\subsection{ File organisation}

To make the metadata reasonably easy for a human to navigate, we
propose splitting the metadata into separate files, all contained
within a \texttt{metadata} folder.  Each \texttt{metadata} will have
exactly one \texttt{dataset.yaml} file and some number of
\texttt{building<I>.yaml} files (where \texttt{I} is an integer).

\subsection{ Example}

\begin{lstlisting}
# dataset.yaml
name: UK-DALE
long_name: >
  UK Domestic Appliance-Level Electricity
meter_devices:
- model: EnviR
  manufacturer: Current Cost
  measurements:
  - physical_quantity: power
    ac_type: apparent
    lower_limit: 0
    upper_limit: 30000

# building1.yaml
instance: 1
rooms:
- {name: kitchen, instance: 1}
- {name: lounge, instance: 1}
elec_meters:

- instance: 1
  device_model: EnviR
  site_meter: true
  sensors:
  - data_location: house1/channel_1.dat

- instance: 2
  device_model: EnviR
  submeter_of: 1
  sensors:
  - data_location: house1/channel_2.dat
  preprocessing:
  - {filter: clip, maximum: 4000}
  appliances:
  - type: light
    components:
    - type: LED lamp
      count: 10
      nominal_consumption: {on_power: 10}
      manufacturer: Philips
      year_of_manufacture: 2011
    - type: dimmer
    on_power_threshold: 10
    main_room_light: true
    dates_active:
    - {start: 2012, end: 2013}
  

\end{lstlisting}

\section{Conclusions}

We have proposed the first draft of a metadata schema for representing
objects relevant to energy disaggregation.  The schema adapts ideas
from DCMI, Project Haystack, ISA\_Tab and DataCite; and adds new
elements relevant to energy disaggregation (only a few of these
elements are required to be instantiated).  We also propose a simple
but powerful inheritance mechanism to minimise duplication of
information and effort. The schema has successfully been used to
capture metadata for the UK-DALE (Domestic Appliance-Level Electricity)
Dataset~\cite{UK-DALE}.

Whilst NILM Metadata is fit for use now, there will inevitably be
use-cases that we have neglected hence we warmly welcome contributions
from the community!  NILM Metadata is open-source to facilitate
collaboration and is available at
\href{https://github.com/nilmtk/nilm_metadata}{github.com/nilmtk/nilm\_metadata}

\section{Future work}

It is currently difficult (if not impossible) to directly compare any
pair of disaggregation algorithms published in the literature.  This
is because different researchers tend to use different datasets and
different performance metrics.  Even if a pair of researchers use the
same dataset, they might use different segments of that dataset!  If
raw disaggregation results and metadata describing the results and
training procedures could be published with each research paper then
the community could begin to objectively rank the performance of
disaggregation algorithms.  Such ranking is common in other fields of
machine learning such as machine vision.  (Ranking energy
disaggregation algorithms is a complex task and we certainly do not
pretend that the introduction of metadata is sufficient!)  As such, we
plan to fully integrate out NILM Metadata schema and object database
with the open source disaggregation toolkit NILMTK~\cite{NILMTK} and
to design a schema for describing disaggregation results using a
combination of our existing `appliance` and `prior` schemas.

Furthermore, NILMTK\cite{NILMTK} already implements the import and
export of appliance models using a simple schema.  We plan to explore
implementing our proposed `appliance models' schema in NILMTK.

\bibliographystyle{plainurl}
\bibliography{library.bib}

\end{document}